\documentstyle[pra,aps,multicol,epsfig]{revtex}

\def\be{\begin{eqnarray}}
\def\ee{\end{eqnarray}}
\def\l{\langle}
\def\r{\rangle}

\begin{document}


\title{Numerical simulations of fundamental processes in cavity QED:
Atomic decay
}
\author{V.Bu\v{z}ek$^{a,b}$, G. Drobn\'y$^{a}$, Min Gyu Kim$^{c}$,
        M. Havukainen$^{d}$, and P.L. Knight$^{e}$
}
\address{
$^{a}$ Institute of Physics, Slovak Academy of Sciences,
D\'{u}bravsk\'{a} cesta 9, 842 28 Bratislava,  Slovakia\\
$^{b}$ Faculty of Mathematics and Physics,
Comenius University, Mlynsk\'{a} dolina F2, Bratislava, Slovakia\\
$^{c}$ Physics Department, Sogang University,
CPO Box 1142, Seoul, 100-611, South Korea \\
$^{d}$ Helsinki Institute of Physics,
P. O. Box 9, FIN-00014 University of Helsinki, Finland\\
$^{e}$ Optics Section, The Blackett Laboratory, Imperial College, London
SW7 2BZ, United Kingdom
}

\maketitle

\begin{abstract}
We present results of numerical investigation of a microscopic
dynamics of a two-level
atom embedded into a ``linear crystal'' of other two-level atoms.
These additional atoms play a r\^ole of a material media.
All atoms interact with a multimode cavity field. We study how the decay of
the initially excited atom is affected by the presence of material media
and spectral properties of the cavity field.
\end{abstract}

\begin{multicols}{2}

\section{Introduction}
\label{sec1}
Quantum electrodynamics (QED) lies at the heart of modern quantum theory.
QED is a well established and experimentally confirmed theory
\cite{Cohen92,Milonni91}  but
even fifty years after its foundation many features of
the atom--field interaction remain extremely intriguing.
In particular, the character of the atom--field interaction
can be substantially modified in {\it confined spaces}
(e.g. the high-$Q$ cavity of a micromaser)
due to the fact that local properties of the electromagnetic (EM)
field depend on space boundaries.
The radiative properties of atoms and the EM field
in confined spaces have been investigated for various cavity
QED systems
\cite{Drexhage74,Kleppner81,Hinds85,Haroche89,Meystre92,Meschede92,Berman94}.

Quantum electrodynamics is a local theory, which means that the dynamics
of atoms and electrons depend on local properties of the electromagnetic
field with which they interact.
But local properties of the electromagnetic field  depend also on conditions
imposed by the boundaries of the space region in which the field is confined
\cite{footnote1}.
These conditions are reflected in the quantization of the field. We can either
quantize the electromagnetic field in free space or in a ``quantization
box'' of linear dimension $L$.
Quantum electrodynamics in a box describes
effects associated with processes inside high-quality (perfect)
cavities.  In addition, quantization in a box can be
considered as an approximation to free-space quantization and
the two theories must give the same results in the limit $L\rightarrow
\infty$.

In the first quantum-mechanical description of	spontaneous decay
of a two-level atom  in  free space,
Weisskopf and Wigner \cite{Weisskopf}  started their calculations
with the cavity modes quantized in a box, and then at a certain
stage of the calculation, a  limit to a continuum of modes was taken.
This approach gives
in first approximation correct results (exponential decay
of the excited level of the atom). The interaction of a two level
atom with discrete (cavity) modes  has been
described systematically by Hamilton \cite{Hamilton} who solved the emission
and scattering problems exactly for a cubic box by
diagonalization of the total atom-field Hamiltonian.
Later this approach was utilized by other researchers
for a detailed investigation of spontaneous emission of two-level
atoms (see for instance papers by
Davidson and Kozak \cite{Davidson} and Swain \cite{Swain}).
Essentially, in all these papers on the spontaneous decay of a two-level
atom in a cavity (box), the coupling constant between
the atom and the cavity modes has been taken to be position independent.
This argument is perfectly justifiable in  free space, when  translational
invariance is valid. On the other hand,
when the atom interacts with discrete
cavity modes in a confined space the position dependence of the
coupling can play a significant r\^ole. The investigation of this problem
is not only of theoretical interest. Recent advances in
experimental techniques have allowed one to study fundamental processes in
 cavity quantum electrodynamics (cavity-QED)
\cite{Drexhage74,Kleppner81,Hinds85,Haroche89,Meystre92,Berman94,Meschede92}
and to verify
various effects of the atom-field interaction in confined spaces as predicted
by Schelkunoff \cite{Schelkunoff}, Purcell \cite{Purcell}, Barton
\cite{Barton} and others.
One of the fundamental processes of cavity QED represents
 the spontaneous decay
of a two--level atom. It is well known that spontaneous emission from
an atom positioned very close to the cavity mirror can be significantly
suppressed. This effect is called the inhibition of spontaneous emission
\cite{Kleppner81}.
The deviation from the exponential Weisskopf--Wigner decay
of the atom in  free space \cite{footnote4}
has been demonstrated in a number of experiments
\cite{Berman94}. Many other interesting questions arise for these QED systems.
For example, one could ask about the influence of cavity mirrors on the
dynamics
of the atom and the r\^ole of the position of the atom
on the appearance of Poincar\'e recurrences
(i.e., re-excitations of the atom by radiation
reflected by the cavity mirrors) \cite{Buzek97}.
While the exponential character of the decay is not affected
by variations of the position of the atom around the center of the cavity
(providing the cavity is large enough), the Poincar\'e recurrences
depend  very sensitively on the position of the atom inside the cavity.
Namely, variation of the position of the atom
within a wavelength of the resonant atomic frequency can result
in an almost complete suppression of the first Poincar\'e recurrence
of the excited level of the atom \cite{Buzek97}.
This means that the atom effectively
does not ``feel'' wave packets reflected from cavity mirrors
 for times much longer than
the time necessary for emitted light to ``travel'' to the mirrors
and back to the atom.
Another example concerns the atom which is positioned close to one
of the mirrors.  In general one may expect to see inhibition
of the radiation. Nevertheless,
taking into account the position dependence of the field--atom interaction
it turns out that for some specific distances from the mirror
(e.g. one quarter of the resonant wavelength of the radiation field),
the atom decays even faster than in  free space \cite{Buzek97}.

A valuable first
insight into modification of the spontaneous emission of the atom
into vacuum field is offered by the Fermi golden rule
\cite{Cohen92}
\be
\Gamma_a = {2 \pi\over{\hbar^2}} |V_{fi}|^2
\rho(\omega_{a})
\label{golden}
.\ee
The spontaneous emission rate $\Gamma_a$ is directly proportional
to the density of field modes $\rho(\omega_{a})$ in frequency domain
at the atomic transition frequency $\omega_{a}$;
$V_{fi}$ is the matrix element of the corresponding transition.
The presence of boundaries
(e.g., in the case when the atom is inserted into a high-$Q$ microwave cavity)
changes the local density of field modes and thereby
the spontaneous emission can be suppressed or enhanced.
However, it is by no means necessary to change the boundary conditions
of the EM field in order to modify the spontaneous emission rate.
This goal can be achieved easily by the presence of other atoms
which can take part in absorption and re-emission of the radiation field.
One important example is that of
an atom embedded in a dielectric host.

The main task of our investigation here concerns  atom-field interactions
in confined geometries. Starting from  ``first principles'' we simulate
the dynamics of a system of atoms in a cavity.
In particular, we consider a cavity filled with a ``crystal''
composed of two--level atoms.
We investigate the modification of the spontaneous emission
and the propagation (scattering) of photon wave packets within
this ``crystal''. We explore the influence of the position
dependence of the atom--field coupling on the dynamics of the system.
In this way also the emission and absorption in
photonic band gap structures (PBS) \cite{Yablonovitch87,PBS,John95}
with few atoms can be analyzed.
Our microscopic model based on ``first principles''
provides us with a deeper understanding of the atom--field interaction
and offers a framework to study systematically
the transition from microscopic to macroscopic (phenomenological)
descriptions of the systems under consideration.

In this paper we focus our attention  on the modifications of spontaneous
emission. We describe the model in Section~\ref{sec2}.
In Section ~\ref{sec3} we study position dependence of the decay
of a single two-level atom and in Section~\ref{sec4} we discuss
the effect of inhibition of spontaneous emission.
We  also analyze how
the dynamics of an initially excited atom is modified
in the presence of other initially deexcited atoms in the cavity
which play the r\^ole of a dielectric (see Section~\ref{sec5}) and
we study in detail
the time evolution of atomic populations and quantum-statistical
properties of the multimode cavity field. In particular,
in Section~\ref{sec6} we focus our
attention on the spectrum of the field. In addition
we will also discuss specific technical questions such as the r\^ole
of the frequency cut-off employed here. In Section~\ref{sec7} we present the
convolutionless master equation describing the dynamics of the initially
excited
atom in dielectrics interacting with multimode cavity field.
We summarize our results in Section~\ref{sec8}.

\section{The model}
\label{sec2}

We consider a simple one-dimensional model of a cavity in which
two-level atoms interact with the cavity modes in the dipole and
the rotating-wave approximations. To simplify
the model, we neglect all mechanical effects of the cavity  field
on the atom (i.e., the mass of the atom  is assumed to be infinite).
Here we note that this 1-D model not only reflects the main features of
atom--field interaction but also can be mapped onto an
isotropic 3-D model.

Under the assumption of perfectly reflecting mirrors,  the operator
of the electric field inside the cavity in the Coulomb gauge
can be expressed as \cite{Cohen92,Stenholm73,Milonni76}
\begin{eqnarray}
\hat{\vec{E}}(r)=\sum_{n,\lambda}{\cal E}_n \vec{\rm e}_\lambda
\left(\hat{a}_{n,\lambda}+\hat{a}^{\dagger}_{n,\lambda}\right) \sin(k_n r),
\label{er}
\end{eqnarray}
where $k_n=\omega_n/c=n \pi/L$ and
${\cal E}_n=\sqrt{\frac{\hbar\omega_n}{\epsilon_0 L}}$.
The two orthogonal polarization vectors
$\vec{\rm e}_{\lambda}$ ($\lambda=1,2$)
lie in the plane perpendicular to the cavity axis;
$\hat{a}_{n,\lambda}$ and $\hat{a}^{\dagger}_{n,\lambda}$ are
annihilation and creation operators of the $n$-th mode.

The Hamiltonian describing the free cavity field can be expressed as
\begin{eqnarray}
\hat{H}_F=\hbar\sum_{\lambda}\sum_{n=1}^N\omega_n
\hat{a}^{\dagger}_{n,\lambda}\hat{a}_{n,\lambda},
\label{hf}
\end{eqnarray}
where we have omitted the zero-point contribution $\hbar\sum_n \omega_n/2$.
Summation over discrete modes in Eq.(\ref{hf}) is performed only up to $n=N$,
which means that in our model we assume a cutoff for the cavity modes.

The Hamiltonian describing a set of $M$ non-interacting (``free'')
two-level atoms with transition frequencies $\omega_a^{(j)}$
can be expressed as
\begin{eqnarray}
\hat{H}_{A}=\frac{\hbar}{2}\sum_{j=1}^M\omega_a^{(j)} \hat{\sigma}_{z}^{(j)},
\label{ha}
\end{eqnarray}
where
$\hat{\sigma}_{z}^{(j)}= |e\rangle_j\langle e| -|g\rangle_j\langle g|$;
$|e\rangle_j$ and $|g\rangle_j$ denote the upper and lower atomic states,
respectively.

When the radius of the atom is much smaller than the wavelength
of the resonant electromagnetic radiation then the atom-field interaction
can be described within the electric-dipole approximation, i.e.,
$\hat{H}_{int}=-\vec{\hat d}\cdot \vec{\hat E}$.
Neglecting for simplicity all polarization effects,
the resulting interaction Hamiltonian in the
rotating-wave approximation (RWA) reads
\begin{eqnarray}
\hat{H}_{int}=
-\hbar\sum_{j=1}^{M}\sum_{n=1}^N
 g_{n}^{(j)} \left[\hat{a}_{n}\hat{\sigma}_+^{(j)}
+\hat{a}^{\dagger}_{n}\hat{\sigma}_-^{(j)}\right],
\label{hint}
\end{eqnarray}
where the Pauli spin-flip operators are
$\hat{\sigma}_{+}^{(j)}= |e\rangle_j\langle g|$ and
$\hat{\sigma}_{-}^{(j)}= |g\rangle_j\langle e|$.
The position dependent coupling constants $ g_{n}^{(j)}$
are given by the expression
\be
 g_{n}^{(j)} =\left(\frac{\omega_n}{\hbar \epsilon_0 L}\right)^{1/2}
{d}_{eg}^{(j)} \sin(k_n r_j)
\label{lambda}
\ee
where ${d}_{eg}^{(j)}$ denotes the dipole matrix elements of the atoms.
The position dependence of the atom-field coupling constant (\ref{lambda})
given by space--mode functions $f_n(r)=\sin(k_n r)$
may significantly affect the atomic dynamics.

The total Hamiltonian of the form
\be
\hat{H}_{tot}=\hat{H}_F+\hat{H}_{A}+ \hat{H}_{int}
\label{htot}
\ee
describes the system of $M$
two-level atoms interacting with $N$ discrete field modes in 1-D cavity.
This model can be solved exactly because
the total number of excitations
\begin{eqnarray}
\hat{R}=\frac{1}{2}\sum_{j=1}^{M}\left(\hat{\sigma}_z^{(j)}+1\right)+
\sum_{n=1}^N \hat{a}^{\dagger}_{n}\hat{a}_{n}.
\label{R}
\end{eqnarray}
is an integral of motion, i.e. $[\hat{R},\hat{H}_{tot}]=0$.

However, it is impossible in general
to find  a closed analytical solution
for the system under consideration. Just a few particular cases can
be solved analytically, and among these is the well known Jaynes-Cummings
model \cite{Jaynes63} which describes the dynamics of a two-level atom
interacting with a {\it single} mode cavity field.
Therefore from the very beginning our treatment
will be based on a numerical simulation of the cavity QED system.
Our numerical treatment allows us
to calculate properties of individual atoms and modes of the EM field,
i.e., it retains a complete microscopic picture of the problem.
The Schr\"{o}dinger equation for the Hamiltonian (\ref{htot}) can be
transformed into a set of coupled linear differential equations for the
amplitudes of component states (e.g., basis of eigenstates of bare
systems) in a finite--dimensional subspace of the Hilbert space.
Numerical solutions can be found using standard methods,
e.g. by Runge--Kutta methods. Alternatively one can apply the
direct diagonalization of the Hamiltonian (\ref{htot}).
Numerical solution allows us
to investigate processes with low initial excitation numbers
(e.g. initially just one or two atoms are excited). The number
 of cavity modes can in this case be
thousands and the total number of atoms can be up to hundreds.

Using  numerical methods, we can analyze
the time evolution of the mean values of the following observables:
\newline
{\bf i)} The occupation of the upper level of the $j$-th atom
\be
\hat{P}_e^{(j)}=\frac{\hat{\sigma}_z^{(j)}+1}{2}=|e\rangle_j\langle e|.
\ee
{\bf ii)} The amplitude of the electric field
\be
\hat{E}(r)=\sum_{n=1}^N\left(\frac{\hbar \omega_n}{\epsilon_0 L}\right)^{1/2}
\left[\hat{a}_n + \hat{a}^{\dagger}_n\right] \sin(k_n r).
\ee
{\bf iii)} The number of excitations of the $k$-th mode
\be
\hat{S}(k)=  \hat{a}^{\dagger}_k\hat{a}_k,
\label{spec1}
\ee
which is used to study the spectrum of the radiation.
\newline
{\bf iv)} To analyze the space-time propagation of  radiation
wave packets we evaluate mean values of the normally-ordered
operator for the energy density
\begin{eqnarray}
\hat{I}(r)= \, :\epsilon_0 \hat{E}^2(r):\,\, .
\end{eqnarray}
Here  normal ordering (the colons above)
is adopted to eliminate the vacuum-state
contribution to the energy density of the emitted radiation.

In what follows we demonstrate the main features of the atom-field
interaction in confined geometries and describe these effects:
\newline
{\bf a)} Modification of spontaneous emission of the atom in the cavity
due to the position dependence of the atom-field interaction.
A partial re-excitation of the atom caused by the back reflected radiation
(Poincar\'e recurrences).
\newline
{\bf b)} Decay in a ``crystal'': modification of spontaneous
emission due to the presence of other atoms, which are initially
deexcited (the decaying atom can be considered as being embedded
in a dielectric ``crystal'' which is formed by other atoms).
\newline
{\bf c)} A  model of quantum measurement: a system of two--level
atoms serves as a device to measure the field spectrum of the considered
configuration. The transient character of the spectrum can be observed.

\section{Decay and re-excitation of atom}
\label{sec3}

Within the Weisskopf--Wigner theory \cite{Weisskopf}
in  free space,
an initially excited atom which is coupled to a continuum of
{\it vacuum} field modes
decays exponentially to its ground state.
Representing the usual
1D continuum with the discrete model (\ref{hint}) for
a large cavity ($L\to\infty$), the population of the excited 
atomic level $P_e$ decays 
exponentially with the rate $\Gamma_a$ given by the Fermi golden
rule (\ref{golden}), i.e.,
\be
P_e(t)=\exp(-\Gamma_a t), \qquad 
\Gamma_a=\frac{\omega_a |d_{eg}|^2}{\epsilon_0\hbar c} 
\label{decay1}
.\ee
In 1D ``free'' space ($L\to\infty$) the decay process is accompanied
by the emission of two wave packets (representing the one-photon state)
propagating to the left and
to the right from the atom. In the case of the ``left-right'' symmetry of
atomic-wave functions in 1-D  
(this corresponds to spherical symmetry in 3-D)
each of the two wave packets carries half of the initial excitation.
This process is irreversible as the energy cannot be reabsorbed by the atom.
In confined geometries the situation differs. First, the density of
{\it discrete} modes is changed due to the boundary conditions.
The translational symmetry is lost and the coupling is position-dependent.
In particular, when the atom is positioned at the center of the
cavity it is coupled only to odd modes of the field (for even modes
the coupling constant (\ref{lambda}) is equal to zero; for more details
see below).
Second, the two wave packets are reflected back by the cavity
mirrors and can be (partially) reabsorbed by the atom.
This partial restoration of the initial state of the atom,
the so called Poincar\'e recurrence, can be viewed as a consequence of
constructive quantum interference (see below).

Figure~\ref{fig1}
presents the time evolution of the probability of
the atomic excitation for four different values of the position of the
atom around the center of the cavity, namely
$\Delta r_1\equiv r_1 -{L\over 2}=
0,\pm{\lambda_a\over 16}\pm{\lambda_a\over 8},\pm{\lambda_a\over 4}$.
To the case when the initially excited atom is positioned 
in the cavity center (dotted line) we will further refer as 
the ``free-space'' decay. 
The central atom interacts only with the odd modes and thus the density of 
(interacting) modes equals to $\frac{L}{2\pi c}$. 
The corresponding ``free-space'' decay rate $\Gamma_a$
is given by Eq.(\ref{decay1}). 

\begin{figure}
\leftline{\epsfig{width=9cm,file=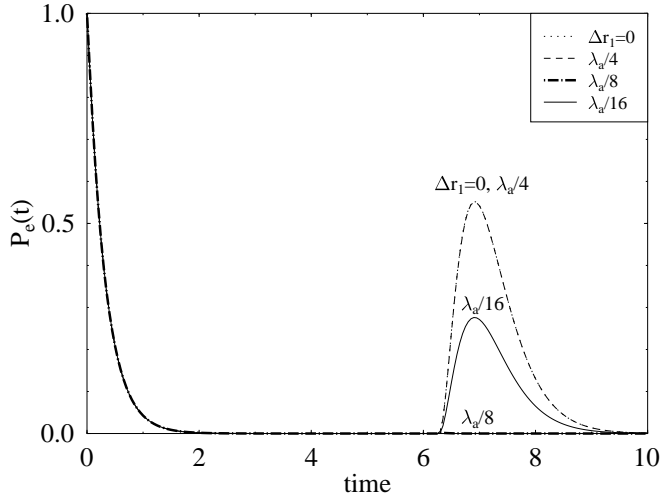}}
\caption{\narrowtext
The time evolution of the population $P_e(t)$
of the excited atomic level. The atom is
shifted from the cavity center by $\Delta r_1=0$ (dotted line),
$\Delta r_1=\pm {\lambda_a/ 4}$ (dashed line),
$\Delta r_1=\pm {\lambda_a/ 8}$ (dashed--dotted line), and
$\Delta r_1=\pm {\lambda_a/ 16}$ (solid line).
The atom is initially prepared in its excited state and
the multimode cavity field is in the vacuum.
The choice of
the cavity length (in dimensionless units) $L=2\pi$,
the squared coupling constant with mode-space function $g_a^2=1/2$ 
(for space-mode function equal to unity) and
the atomic transition frequency $\omega_a=100$
leads to $\Gamma_a=\pi$ and $\lambda_a=L/50$.
The first Poincar\'e recurrences appear at the time $t_R=2\pi$.
The upper cutoff on frequencies is set to $\omega_{cut}=200$.
}
\label{fig1}
\end{figure}

~From Fig.~\ref{fig1} we see that the first ``exponential'' stage of the 
decay is (almost) position independent.
Providing the atom is ``far'' from the mirrors, i.e., 
$\min(r_1,L-r_1) \gg c/\Gamma_a$, the reflected wave packets 
do not influence the exponential decay.
More precisely, the density of modes is doubled when the atom is shifted 
from the cavity center. Owing to the position dependence of the
atom-field coupling (\ref{lambda}) the even modes start to interact 
with the atom and the interaction with the odd modes decreases.
Considering the effective squared interaction constant
as the average for two neighboring (odd and even) cavity modes 
it decreases to the half value of the squared coupling 
constant of the interacting modes for the atom in the cavity center.
It means that even though the atom is shifted from the cavity center
the Fermi golden rule (\ref{golden}) with the effective squared interaction
constant and doubled density of modes leads to the same decay rate 
(\ref{decay1}).

For large enough times, the total excitation energy of the atom is
transferred to the field, which in turn is effectively in a one photon
(one excitation) state represented by  two EM wave packets propagating
towards the mirrors. For finite cavities at time approximately
${L\over 2c}$ the wave packets are reflected by the mirrors and
at $t_R \simeq {L\over c}$ they approach the atom, which starts to reabsorb
the energy from the field. We observe the re-excitation of the atom
(i.e. the Poincar\'e recurrence can be observed).
In contrast  to the ``exponential'' stage of the atomic decay,
Poincar\'e recurrences are very sensitive to small position shifts
of the atom within a wavelength of the resonant atomic
transition. In Fig.~\ref{fig1} we clearly see that if the atom is positioned
at the cavity center ($r_1=L/2$) then at time $t_R\simeq L/c$
a very well pronounced Poincar\'e recurrence of the atomic
inversion is seen.
One can say that at the moment when the Poincar\'e recurrence
appears the atom ``sees'' the cavity mirrors \cite{Parker87}.
On the other hand, with a small shift of the atom from the cavity
center to $\Delta r_1=\pm {\lambda_a \over 8}$
the first atomic recurrence  is almost completely suppressed.
For simplicity we consider the two
emitted wave packets (one to the left and one to the right)  as monochromatic
plane waves (at the atomic transition frequency and with the group velocity
$c$). The difference of their geometrical paths is equal to
${\lambda_a\over 2}$. This path difference results in
destructive interference due to the accumulated phase difference of $\pi$.
In other words, the atom does not ``see'' the wave packets reflected from the
cavity mirrors. Obviously, when the two wave packets propagate further, then
after the second reflection they accumulate
 a phase difference of  $2\pi$ so the corresponding
Poincar\'e recurrence can then be seen
(i.e. in this case
the atom needs an elapsed time which is twice as long compared with the
situation
when $r_1={L\over 2}$  to ``see'' the cavity mirrors).

For the case of an atom positioned at
$\Delta r_1=\pm{\lambda_a\over 4}$ the evolution of the atomic inversion
is almost indistinguishable from the case where $\Delta r_1=0$.
At the time of appearance of the first Poincar\'e recurrence there is
a constructive interference of the wave packets.
The trivial phase shift $2\pi$ results from the difference of
the geometrical paths which is then equal to $\lambda_a$.

In the case of an atom positioned at $\Delta r_1=\pm{\lambda_a\over 16}$
the path difference is equal to $\lambda_a/4$ and the first Poincar\'e
recurrence is intermediate between the extreme
cases ($\Delta r_1=0,\pm{\lambda_a\over 8}$) considered above.
Dephasing of the wave packets by ${\pi\over 2}$
results in a partially reduced reabsorption.
More rigorous analysis should take into account
the multimode structure of the wave packets as an additional source of
dephasing due to the different (Rabi) frequencies of the modes.
We next note that the second Poincar\'e recurrence associated with the
second reflection from the boundaries starts simultaneously
in all cases shown in Fig.~\ref{fig1}.
At the time $\simeq 2 t_R$ the wave packets are merging
in-phase, i.e. their  geometrical paths are equal, which results in
a partial reexcitation of the atom.

\begin{figure}
\leftline{\epsfig{width=9cm,file=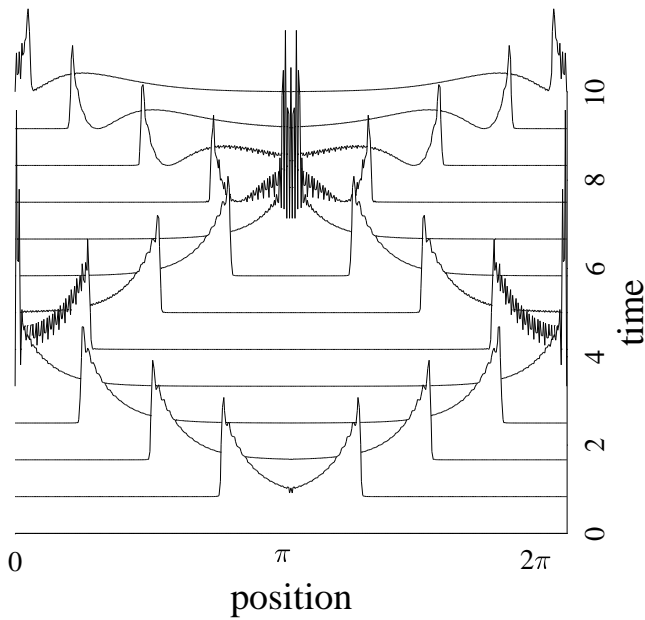}}
\caption{\narrowtext
A stroboscopic set of plots describing the
space-time propagation of the mean energy density of the cavity field.
We assume the same settings as in  Fig.~\ref{fig1} with
the atom located in the cavity center.
We see the two wave packets propagating towards the cavity mirrors. These
wave packets have ``sharp'' fronts.
 We note that larger the number of modes
coupled to the atom the sharper the fronts are.
The length of the   tails of the wave packets
depends on the life-time of the atom.
}
\label{fig1b}
\end{figure}

Summarizing this part of the description of the dynamics of the atom,
we can say  that while the ``exponential'' character of the
decay of an excited atom inside a large cavity is not influenced by
small shifts of the atomic position, the first Poincar\'e recurrence
is a position-dependent interference effect.
The basic features can be explained using a simple {\it classical}
point of view based on two interfering monochromatic waves.
This fact may be thought of as rather surprising, as mathematically
the Poincar\'e recurrences can be related to the discrete
nature of the cavity modes (with equal frequency spacing).
To be more precise, the phase-matching conditions necessary for
the appearance of Poincar\'e recurrences in the atom-field system
can be associated with the evolving phase factors $\mbox{e}^{-itE_k}$
of contributing eigenstates $|\Phi_k\rangle$ of the total Hamiltonian
(\ref{htot}):
a Poincar\'e recurrence, i.e., a partial restoration of the initial state,
can appear at time $t_R$ such that the relation
$E_k t_R \simeq 2\pi$ is valid for many values of $k$
(for more details see \cite{Milonni83}).

In Fig.~\ref{fig1b} we present a stroboscopic set of plots describing the
space-time propagation of the energy density
of the cavity field for the same physical situation (initial state)
as considered in Fig.~\ref{fig1} when the atom is in the center of the
cavity.
We see two distinct wave packets propagating to the right
and to the left. Reflection of the wave packets from the
cavity mirrors (at time $t\simeq {L\over 2c}$) is nicely demonstrated
and the subsequent re-excitation of the atom is synchronized with the
interference of the wave packets in the center of the cavity (compare
with Fig.~\ref{fig1}).

\section{Inhibition of spontaneous emission}
\label{sec4}

\begin{figure}
\leftline{\epsfig{width=9cm,file=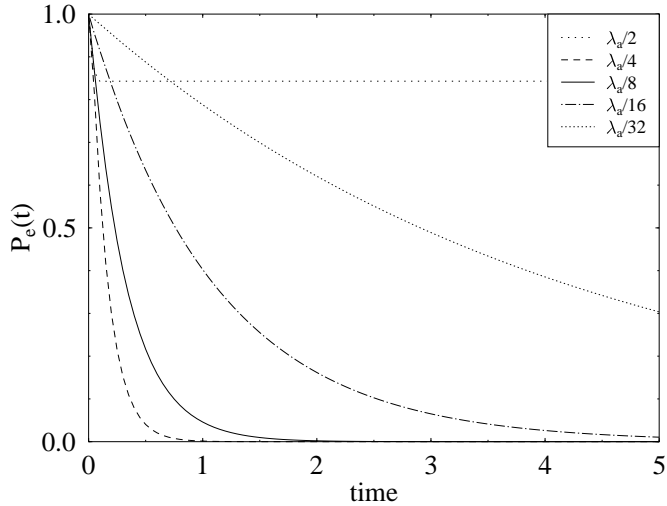}}
\caption{\narrowtext
The time evolution of the population $P_e(t)$
of the excited atomic level for the atom very close to the cavity mirror.
The atom is considered at the following positions:
$r_1={\lambda_a/2}$ (dotted line), $r_1={\lambda_a/4}$ (dashed line),
$r_1={\lambda_a/8}$ (solid line), $r_1={\lambda_a/16}$ (dashed-dotted line)
and $r_1={\lambda_a/32}$ (dotted line).
The ``reference'' exponential
decay of the atom at the cavity center $r_1=L/2$
coincides with the case $r_1=\lambda_a/8$.
The initial conditions and other parameters are as in Fig.~\ref{fig1}.
}
\label{fig2}
\end{figure}

In the previous section we have considered situations when the atom is
``far'' from
the cavity mirrors [i.e. $\min(r_1,L-r_1) \gg c/\Gamma_a$] and
the wave packets reflected by the mirrors  do not directly affect
the initial spontaneous decay of the atom.
On the other hand, for distances between the atom and one of the cavity
mirrors smaller than $c/\Gamma_a$ (here $1/\Gamma_a$ is
the spontaneous decay time in a free space) deviations
from exponential decay should be expected \cite{Kleppner81,Hinds85,Haroche89}.
In particular, the decay of a two-level atom which is positioned very
close to the cavity mirror can be significantly suppressed.
This effect is called the inhibition of spontaneous emission
\cite{Kleppner81}.
The inhibition of spontaneous emission
is a position-dependent effect which is related to
the position dependence of the atom-field coupling constant (\ref{lambda}).
In Fig.~\ref{fig2} we present numerical simulations
for the time evolution of the population of the
upper level of the atom described by
the model interaction Hamiltonian (\ref{hint}).
The atom  is assumed to be initially
in its excited state and the field in the vacuum.
We consider several typical physical configurations. Firstly,
for reference, we plot
the atomic decay of the atom positioned at
$r_1=\lambda_a/8$ (solid line)
which is indistinguishable from the exponential decay of the atom
at the cavity center (i.e., $P_e\approx\mbox{e}^{-\Gamma_a t}$
for $t\leq t_R$).
For other atomic positions $r_1=\lambda_a/16$ and $r_1=\lambda_a/32$
(here $\lambda_a=L/50$)
we clearly see that the closer the atom is to the mirror
the slower the spontaneous decay is, that is the inhibition of spontaneous
radiation is transparent for the considered positions of the atom.
On the other hand  for very specific atomic
positions close to the mirror the opposite effect -
the {\em enhancement} of spontaneous emission - takes place.
Namely, for the atomic distance $r_1=\lambda_a/4$
the atom decays as $P_e\approx\mbox{e}^{- 2 \Gamma_a t}$,
i.e., it radiates twice as fast compared with the free-space case
[see the reference case $r_1=\lambda_a/8$].

The origin of inhibition or enhancement of spontaneous
emission in the context of the model used in this paper
relies on the position dependence of the atom-field coupling
(\ref{lambda}). In particular, for $r_1=\lambda_a/4$
the spatial--mode function $\sin(k_n r_1)\approx 1$
for all the modes close to the resonant frequency $\omega_a$
irrespective of whether $n$ is even or odd.
This means that the density of modes is increased by a factor of two
 compared with
the case of the atom at the cavity center $r_1=L/2$ when the modes with
even $n$ are decoupled from the atom [$\sin(k_n L/2)=0$ for even $n$
and $\sin(k_n L/2)=1$ for odd $n$]. The increased density of modes
implies an enhancement of the spontaneous emission.
In a similar way, when $r_1=\lambda_a/8$
the spatial--mode function $\sin(k_n r_1)\approx 1/\sqrt{2}$
for all $n$ around the atomic transition frequency.
However, the decrease of the squared
interaction constants is compensated for by an increase in the density of
interacting modes (compared with the atom at the cavity center)
and thus the spontaneous emission rate retains the value $\Gamma_a$.
For the other extreme case $r_1=\lambda_a/2$ all modes around $\omega_a$
are significantly decoupled from the atom [now $\sin(k_n r_1)\approx 0$]
which results in the dramatic inhibition of the spontaneous emission
as seen in Fig.~\ref{fig2}.
For other positions shown in Fig.~\ref{fig2} ($r_1=\lambda_a/16,\lambda_a/32$)
the slowing of the spontaneous emission rate is given by the
factor $\Gamma(r_1)/\Gamma_a \approx 1-\cos(2 k_a r_1)$.
We note that in the case $r_1=\lambda_a/16$ the atom decays
completely while for $r_1 = \lambda_a/32$ the exponential decay law
is interrupted by the Poincar\'e recurrence at $2t_R$.

\begin{figure}
\leftline{\epsfig{width=9cm,file=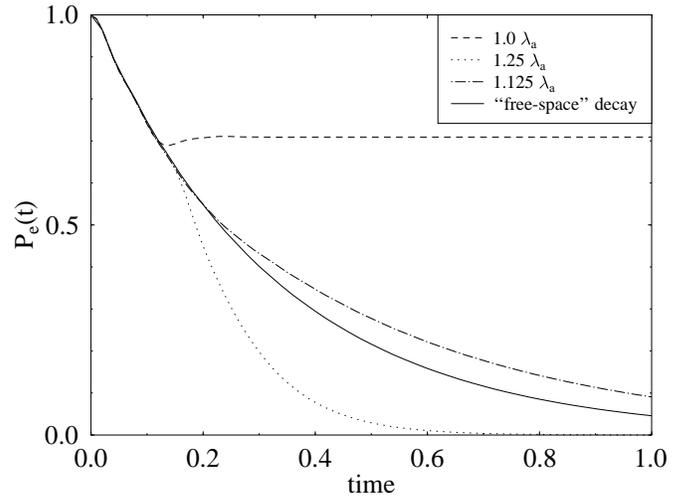}}
\caption{\narrowtext
The time evolution of the population $P_e(t)$
of the excited atomic level for the atom
 at the following positions:
 $r_1=\lambda_a$ (dashed line);
 $r_1=\lambda_a +\lambda_a/4$ (dotted line);
and $r_1=\lambda_a+\lambda_a/8$ (dashed-dotted line) which are compared
with the exponential decay of the atom in the cavity center (solid line).
Other settings are as in  Fig.~\ref{fig1}.
The suppression and the stimulation of the emission caused by the
reflected wave packet are clearly see.
}
\label{fig3}
\end{figure}

For  distances of the atom  from the
mirror larger than the wavelength of the resonant transition $\lambda_a$
the interference with  the reflected wave packet can either
 stimulate or suppress the emission of the atom. To be specific
we show in Fig.~\ref{fig3} the time evolution of the population of the upper
level of the atom which is positioned at three  distances
 $r_1=\lambda_a$ (dashed line);
 $r_1=\lambda_a +\lambda_a/4$ (dotted line);
and $r_1=\lambda_a+\lambda_a/8$ (dashed-dotted line) which are compared
with the exponential decay of the atom in the cavity center (solid line).
The phase accumulated by the wave packet during
 the round trip from the atom
 to the neighboring mirror and back is in the case $r_1=\lambda_a$
equal approximately to $5\pi$
(here the additional contribution of $\pi$ is
due to the reflection from the mirror), i.e. there is a destructive
interference between the wave packet and the atom which results in
the suppression of  the radiation. On the other hand, when
$r_1=\lambda_a +\lambda_a/4$ the accumulated phase is approximately
$6\pi$, which leads to constructive interference.
It means that the reflected wave packet, when it arrives at the
position of the atom, starts to stimulate the atomic emission.
In the units used in this simulation, the arrival time of the reflected
wave packet is at approximately  $t\simeq 0.16$ which coincides with the
change of the initial exponential decay of the atom.
When the atom is at the position
$r_1=\lambda_a+\lambda_a/8$, the accumulated phase of the reflected wave
packet is $11\pi/2$ which gives rise to a partial suppression of radiation.

\section{Spontaneous emission in a ``crystal''}
\label{sec5}

Atomic radiation can be crucially modified by the presence of other
atoms in the cavity. Obviously, if the distance between the atoms
is large enough then the  exponential decay of the originally excited atom
is not affected significantly. On the other hand when the atoms are placed
close together the situation is different (one of the consequences
is a collective behavior of the atoms which might result in
superradiance, see for instance \cite{Dicke}).

In this section we
consider a specific initial condition when the initially excited atom
is surrounded by a collection of two-level atoms in the ground state.
These additional atoms    can be considered as  a linear ``crystal''.
By changing the density of the atoms we can model systems
such as atomic structures embedded in optical lattices
(for interatomic distances comparable with the wavelength
of the atomic transition) or dielectrics
(for much smaller interatomic distances).
The cavity QED system with trapped atoms \cite{Buzek97b}
represents a quite new experimental system.
The transfer of excitations between particular atoms which are captured
in an optical potential is mediated by the cavity field \cite{foot6}.

\begin{figure}
\leftline{\epsfig{width=9cm,file=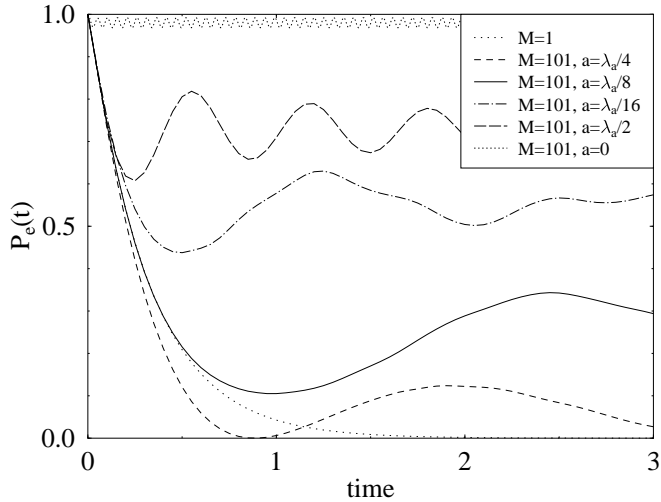}}
\caption{\narrowtext
The modification of the spontaneous emission of the atom
at the cavity center being surrounded by identical atoms which
form a linear ``crystal''.
The central atom is initially excited and the others de-excited.
The vacuum field is in the vacuum.
The regular linear crystal which fills the central part of the cavity
is composed of $M=101$ atoms with the interatomic distance
$a=\lambda_a/2$ (long dashed line), $a=\lambda_a/4$ (short dashed line),
$a=\lambda_a/8$ (solid line), $a=\lambda_a/16$ (dot--dashed line).
In the case $a=0$ (dotted line) all the atoms are positioned
at the cavity center. Single atom decay $M=1$ (sparse dotted line)
is shown for reference.
}
\label{fig4}
\end{figure}

The modification of the spontaneous emission of the atom
embedded in a linear ``crystal'' of two--level atoms is shown in
Fig.~\ref{fig4}.
The regular crystal built of $M=101$ atoms fills the central part
of the cavity. Initially the only excited atom
is located at the cavity center and decays in the vacuum field.
The modification of the spontaneous emission
depends dramatically on the interatomic distance $a$.
In the case of the ``lattice'' constant being $a=\lambda_a/2$
(long dashed line) we can observe strong suppression of the spontaneous
emission while for $a=\lambda_a/4$ (short dashed line) an enhancement
of radiation
compared with the single atom system (dotted line) takes place.
The origin of this behavior is related to either destructive or
constructive interference effects, respectively. From other examples,
for $a=\lambda_a/8$ (solid line) and
$a=\lambda_a/16$ (dot--dashed line) it is seen that by
increasing the density of the linear ``crystal'' the atom radiates
more slowly. Moreover, the de-excitation is incomplete, as
an increasing part of the excitation is captured by the initial state.
This subradiant behavior has been  already analyzed for the extreme case when
all the atoms are located at the same position (e.g., the cavity center)
\cite{Buzek89}. The initial excitation is captured in the asymmetric atomic
state and only a small part $\sim \frac{1}{M}$ is radiated in
the cavity field.

\begin{figure}
\leftline{\epsfig{width=9cm,file=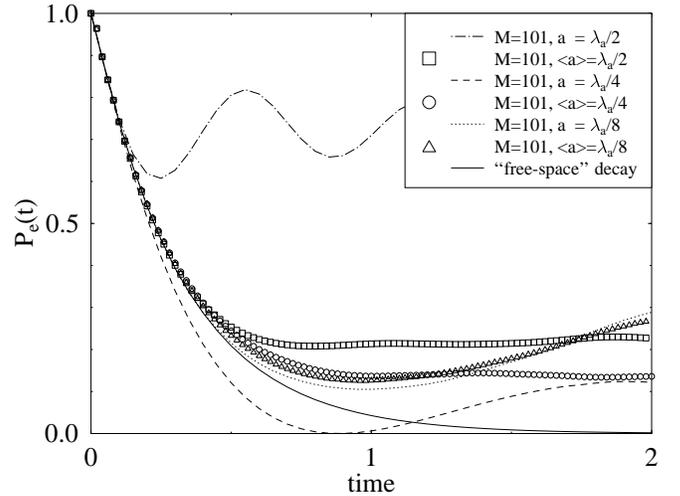}}
\caption{\narrowtext
The time evolution of the population $P_e(t)$
of the excited atomic level for the atom surrounded
by identical atoms which form a random linear structure.
Averages over 100 random configurations
with one atom within the lattice constant
$a=\l\lambda_a/8\r$ (triangles),
$a=\l\lambda_a/4\r$ (circles), and
$a=\l\lambda_a/2\r$ (squares)
are compared with the corresponding regular crystals
for $a=\lambda_a/8$ (dotted line), $a=\lambda_a/4$ (dashed line),
and $a=\lambda_a/2$ (dot--dashed line). 
Setting are as in Fig.~\ref{fig4}.
Single atom decay ($M=1$) in ``free-space'' (solid line)
is shown for reference.
}
\label{fig5}
\end{figure}

The regular ``crystal'' represents a rather specific and idealized case.
Positions of atoms can fluctuate due to various reasons
(for example in the case of optical lattices with shallow wells formed from
optical potentials).
To simulate the situation when the  atoms  are not  regularly
distributed in the cavity we next consider random configurations
of the atoms. Specifically, the atoms are placed randomly such that
within each lattice constant there is just one atom.
Depending on the particular positions of the atoms,
the  dynamics of the originally
excited atom can change dramatically. The atomic radiation can be either
enhanced or suppressed. To obtain some effective ``macroscopic'' picture
from our simulations, we have averaged over many
random configurations.

We present the results in Fig.~\ref{fig5}.
The dashed (dotted) line in this figure shows
the time evolution of the atomic population of the initially excited atom
when the atoms are regularly positioned with the lattice constant
$a=\lambda_a/4$ ($a=\lambda_a/8$), representing enhancement (suppression)
of radiation with respect to the ``free-space'' decay (solid line).
The results of numerical simulations corresponding to averaging over
many (100) random configurations of atoms are presented
for the average distance between atoms
$\l a\r=\lambda_a/4$ ($\circ$) and $\l a\r=\lambda_a/8$ ($\triangle$).
In both cases the radiation of the atom is suppressed compared with the
``free-space'' decay. Another common feature of the dynamics in this case
is that in both cases the atom does not radiate away completely the initial
excitation energy. Both effects (suppression of radiation and ``excitation
trapping'') are caused by the collective influence of the crystal atoms.

An increase of the density of the linear ``crystal''
(e.g. for $a=\lambda_a/16$) rapidly diminishes differences
between the regular crystal and the corresponding ``random'' crystal
with one atom randomly positioned within the lattice constant.
For completeness we included in Fig.\ref{fig5} also the case of
the regular ``crystal'' with $a=\lambda_a/2$ (\fbox{})
and the average over random configurations with $\l a\r=\lambda_a/2$.
Here the destructive interference effect
which leads to the strong inhibition of the radiation
for the regular crystal is deteriorated for random atomic configurations
shown in Fig.\ref{fig5}.

\begin{figure}
\leftline{\epsfig{width=9cm,file=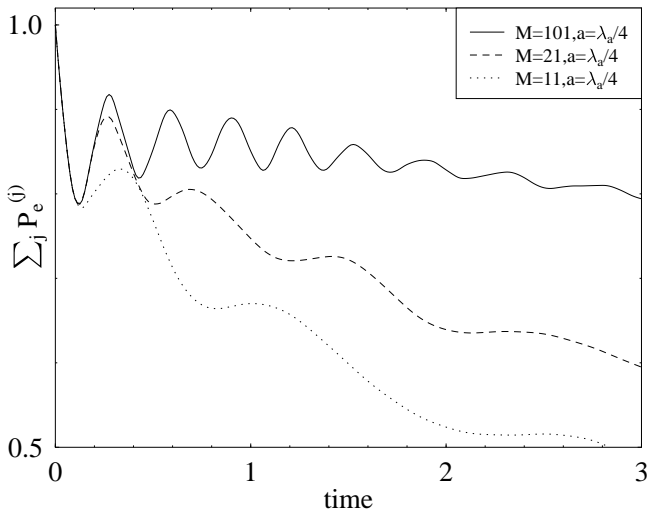}}
\caption{\narrowtext
The total excitation of the atoms $R_{atoms}=\sum P_e^{(j)}$
which form a regular linear ``crystal'' as in Fig.~\ref{fig4}.
For the interatomic distance $a=\lambda_a/4$ also small ``crystals''
made of $M=11$ (dotted line) and $M=21$ (dot--dashed line) are
considered.
}
\label{fig6}
\end{figure}

We note that
the modification of the spontaneous emission is a local effect,
i.e., the atomic decay is influenced only by neighboring atoms.
To check this we have performed simulations
with only 10 neighboring atoms ($M=11$). We have found that the
 ``exponential'' stage of the spontaneous emission is unchanged
comparing with the case $M=101$, and differences occur only on a much
longer time scale.

It is instructive to note that part of the initial excitation
energy is captured in the atomic system.
In particular, the sum of atomic excitations
$R_{atoms}=\sum P_e^{(j)}(t)$ depends mainly
on the number of atoms $M$. This tendency is confirmed in Fig.~\ref{fig6}
which shows the cases $M=11,21,101$ for $a=\lambda_a/4$.
In general, the oscillation patterns reflect complex interference effects.
Nevertheless one can trace a very general tendency in the picture - the
``crystal'' atoms which surround the initially excited atom play the r\^ole
of semi-transparent mirrors  placed very close to the atom.
Therefore the results partially resemble the case of the single atom
in the vicinity of a mirror (compare Fig.~\ref{fig4} with Fig.~\ref{fig2}).

\section{Spectrum of the cavity field}
\label{sec6}

Within the framework of cavity QED when the field interacting with the atoms
is confined within ideal mirrors, there is nothing like a stationary regime
which is necessary for the derivation of a time-independent
spectrum of the field. The spectrum is intrinsically time dependent.
In this case an operational definition of time-dependent spectrum can
be given by excitation probabilities of the cavity modes
[see Eq.(\ref{spec1})].

The spectrum of the cavity field is affected by the position
of the atom. In particular, if the atom is located in the cavity center
even modes are completely decoupled from the atom and only odd modes can
become excited [see Eq.(\ref{lambda})]
establishing in this way  oscillations in the spectrum of modes.
However, at the point when the total excitation energy of the
atom is transferred to the field, the {\em envelope} of the spectrum
is ``Lorentzian'' irrespective of the position of the atom
(providing that the decay is exponential).

\begin{figure}
\centerline{\epsfig{width=15cm,file=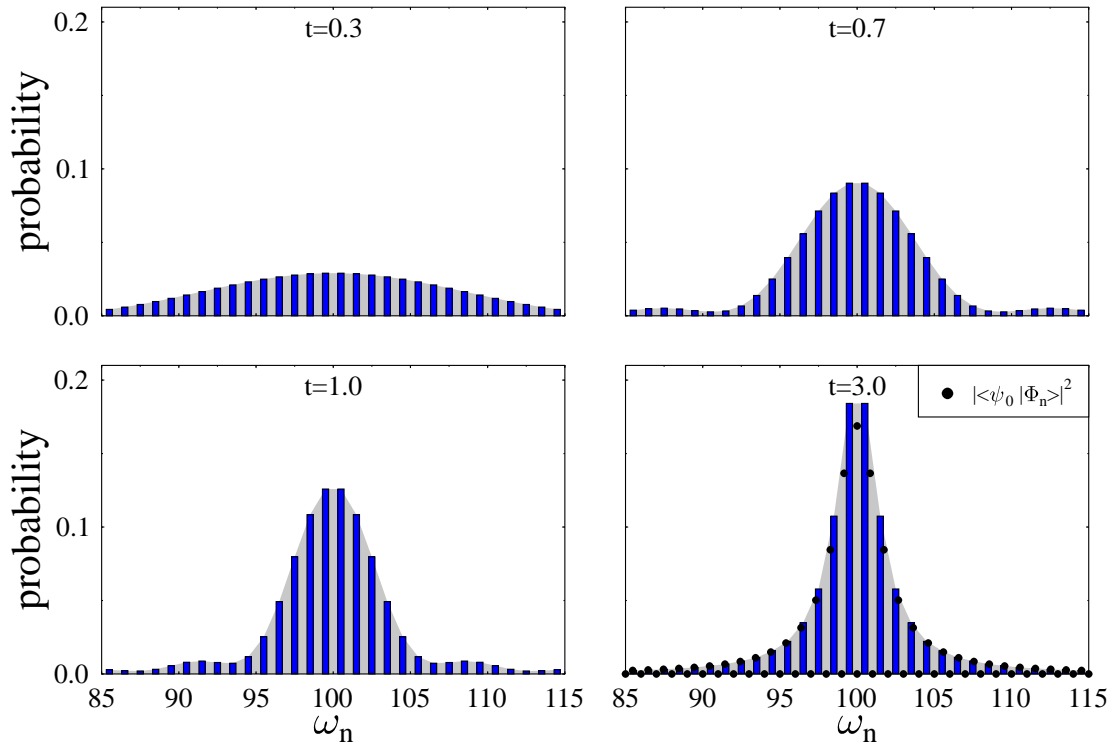}}
\caption{
Populations of cavity modes (i.e. the spectrum)
at times $t=0.3,0.7,1,3$ (in units of $1/\Gamma_a$)
for the atom located at the cavity center.
The atom is initially prepared in its excited state and
the cavity field is in the vacuum (for other conditions see Fig.~\ref{fig1}).
For comparison purposes we show
the overlaps of the eigenstates of the total Hamiltonian (\ref{htot})
with the initial state
(these overlaps are denoted by the symbol $\Delta$).
The ordering of eigenstates is given by their eigenvalues
on frequency axis. The even modes which do not interact with the atom
have no overlap with eigenstates of the total Hamiltonian 
(see oscillations in the function $|\l\psi_0|\Phi_k \r|^2$).
}
\label{fig7}
\end{figure}

On the other hand, it should be stressed that the spectrum
of the interacting modes is highly transient even during
the exponential decay period.
It undergoes a gradual narrowing from a broad flat spectrum
(initially all modes are in the vacuum state with the same probability)
towards a  Lorentzian-like line of  width $\Gamma_a$.
The narrowing is accompanied by
 transient oscillations of the
spectral envelope. This transient behavior is illustrated in Fig.~\ref{fig7}
which shows the spectral envelopes at different time moments during
the exponential decay of the atomic excitation.
At the time $t\approx 2$ the envelope of the cavity-field spectrum
reaches its quasi-stationary shape, being very close to
the corresponding (Lorentzian) emission spectrum usually associated
with the free-space emission \cite{Cohen92}.

\end{multicols}

\widetext

\begin{multicols}{2}

It is worth noticing that there is a close relation between the
emission spectrum and the ``spectrum'' of squared scalar products
(overlaps)
between eigenvectors $|\Phi_k \r$ of the total Hamiltonian (\ref{htot})
and the given initial state $|\Psi(t_0)\r$, i.e.,
\be
S_e(k)=|\l \Psi(t_0) | \Phi_k \r |^2
.
\ee
\null From Fig.~\ref{fig7} it is evident that the ``spectrum''
of overlaps (shown as $\bullet$) resembles the emission spectrum
of the completely deexcited atom.
In other words, the ``spectrum'' of overlaps offers an important
time-independent characteristization of the system under consideration.
If there exists a quasi-stationary spectrum of the cavity modes,
it should mimic the ``spectrum'' of overlaps.
In addition, a shift of the atomic transition frequency
in the spectrum of eigenvalues can be associated with the
energy shift.
In our calculations we use a  broadband approximation,
in which the frequency dependence of the atom-field coupling
can be neglected. Thus for the symmetrical upper frequency cutoff
$\omega_{cut}=2\omega_a$ the energy shift equals  zero.

We next turn our attention to the fact that an additional system
of two--level atoms with different transition frequencies $\omega_a^{(j)}$
inside the cavity can serve as a device to measure
the spectrum of the cavity modes. Within this  model of measurement
the initially deexcited (analyzer) atoms are  placed far enough from
the  decaying atom. The coupling of the analyzer atoms to the cavity modes
is so weak that the dynamics of the cavity field is essentially
unaffected on the relevant time scale.
It means that  the linewidths of the analyzer
atoms are much smaller than the linewidth $\Gamma_a$ of the decaying
atom. The analyzer atoms with different transition frequencies
$\omega_a^{(j)}$ thus function as narrow frequency filters interacting
effectively only with the cavity modes on exact resonance with
particular $\omega_a^{(j)}$.
The upper level excitation probabilities of the analyzer atoms are
proportional to the intensity of the cavity field at the position
of the analyzer atoms \cite{Martti}. In order to map the excitations
of analyzer atoms to the spectrum of the cavity modes it is necessary
to demand that the coupling constants (linewidths) of the analyzer atoms
are equal.
In general, the analyzer atoms give the local and time dependent
frequency spectrum which is not necessarily the same as 
excitations of cavity modes (\ref{spec1}) which give the ``instantaneous'' 
spectrum in the whole cavity. In the case of the atomic decay 
the two spectra agree if the times they are read out from the analyzer 
atoms and the modes are chosen appropriately.

\begin{figure}
\leftline{\epsfig{width=7cm,angle=90,file=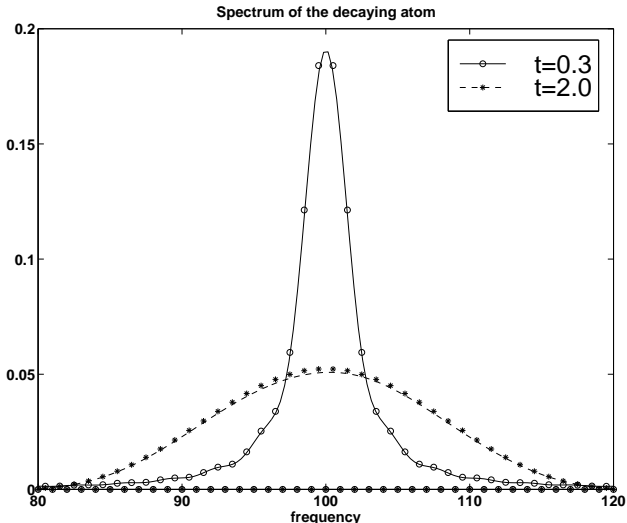}}
\caption{\narrowtext
The spectrum of the cavity modes at times
$t=0.3$ ($\star$) and $t=2$ ($\circ$) is
compared with the normalized absorption spectrum
of the analyzer atoms at times $t=0.3+t_{f}$ (dashed line)
and  $t=2+t_{f}$ (solid line).
Here $t_{f}$ is time of flight of the light
from the decaying atom at the cavity center with $\Gamma_a$ as in
Fig.~\ref{fig1} to the analyzer atoms with mutually equal
decay rates $\Gamma=10^{-4}\Gamma_a$. 
The distance of the analyzer atoms from decaying atoms is $\Delta r=0.5$
and the time of flight $t_f=0.5$ in choosen units ($c=1$).
}
\label{fig7b}
\end{figure}

The normalized absorption spectrum read out
from the analyzer atoms is show in Fig.~\ref{fig7b}.
Here we have considered that the
decay rates (i.e. linewidths) $\Gamma$ of the analyzer atoms
are mutually equal and $\Gamma=10^{-4}\Gamma_a$.
A bunch of one hundred of analyzer atoms is positioned $\Delta r=0.5$
apart of the decaying atom in the cavity center.
The atomic frequencies $\omega_a^{(j)}$ are equally spaced around
the transition frequency $\omega_a$ of the central atom.
We see very good agreement
with the {\em envelope} of the spectrum of the cavity modes.
More precisely, in  Fig.~\ref{fig7b} we compare the spectrum of the cavity
modes at times $t=0.3$ and $t=2$ with the absorption spectrum of the analyzer
atoms at delayed times $t+t_{f}$ where $t_{f}$
is the time of flight of the light from the decaying (central)
atom to the analyzer atoms.

\subsection{Numerics and cutoff dependence}
\label{sec6.c}

Our  numerical calculations have been performed
in the broadband approximation for the interaction constants given by
 Eq.(\ref{lambda}), i.e.,
we have neglected the frequency dependence of the coupling constants
replacing $\omega_n$ by $\omega_a$.
This approximation is valid only for large enough cavities
with $L\gg\lambda_a$ and ``weak'' interaction regimes
with $\Gamma_a\ll \omega_a$.

A rather subtle point is the choice of the frequency cutoff.
Strictly speaking, the model interaction Hamiltonian (\ref{hint})
with the interaction constants (\ref{lambda}) within
the broadband approximation leads in second-order perturbation
theory to {\it logarithmically} divergent
energy shifts for $\omega_{cut}\to\infty$ \cite{Milonni76}.
(Note that in our numerical calculations we have eliminated the shift of the
excited atomic level by choosing a symmetrical frequency cutoff
$\omega_{cut}=2\omega_a$.) Without the broadband
approximation, when the frequency dependence of the interaction
constants (\ref{lambda}) is taken into account,
the energy shifts diverge {\it linearly}.
It is well known \cite{Cohen92,Milonni91} that
 if instead of the dipole approximation
$\hat{H}_{int}=-\vec{\hat d}\cdot \vec{\hat E}$ we  start
with $\hat{H}_{int}=-\vec{\hat p}\cdot \vec{\hat A}$
then after the RWA is applied
 one obtains   the interaction Hamiltonian (\ref{hint})
but with a different frequency dependence of the interaction constant, i.e.
\be
 g_{n}^{(j)} =
\sqrt{\frac{\omega_a}{\omega_n}} \left(\frac{\omega_a}{\hbar\epsilon_0
L}\right)^{1/2}
{d}_{eg}^{(j)}\sin(k_n r_j)
.\label{lambda2}
\ee
In the broadband approximation the interaction constants
given by Eq.(\ref{lambda}) and Eq.(\ref{lambda2}) are identical and
the results do not depend on the choice of the interaction Hamiltonian.
On the other hand, without the broadband approximation the results are
biased by the choice of the frequency dependence of the atom--field
coupling. From the mathematical point of view,
the coupling given by the expression (\ref{lambda2}) does not lead
in second-order perturbation theory to divergent energy shifts
for $\omega_{cut}\to\infty$.
Obviously at the point when the two effective Hamiltonians considered
above lead to different results
one has to be careful whether the model is physically relevant
(for more details see Ref.~\cite{Milonni91}).

\section{Master equation for the atom in dielectrics}
\label{sec7}

The system of atoms and field modes under consideration in
an ideal cavity represents a closed system with unitary dynamics
governed by Schr\"{o}dinger equation. In this Section we will consider
the decaying atom as an open system in the environment represented
by field modes and other initially unexcited atoms.
This analysis
can give us a deeper insight into the problem of dynamical evolution
of the atom in dielectrics interacting with many cavity modes.
In general, an open system $S$
(in our case the atom which is initially excited)
interacts with an environment $E$ (the other atoms surrounding the
originally excited atom and cavity modes)
 \cite{Davies}. In this section we consider the archetypal
{\it system $+$ environment} model which is specified as follows:
Let ${\cal H}_{_S}$ denotes a
Hilbert space of the system $S$, and ${\cal H}_{_E}$ is the
Hilbert space associated with the environment $E$.
The Hamiltonian
$
\hat{H}_{_{SE}}= \hat{H}_{_S}\otimes \hat{1}_{_E}
+ \hat{H}_{int} + \hat{1}_{_S}\otimes \hat{H}_{_E}
$
of the composite system $S\oplus E$
acts on ${\cal H}_{_S}\otimes{\cal H}_{_E}$. It is assumed that $S\oplus E$
is a {\it closed finite-dimensional}
 system which evolves unitarily.  The density
operator $\hat{\rho}_{_{SE}}(t)$ of this composite system is governed by the
von Neumann equation with the formal solution $\hat{\rho}_{_{SE}}(t)=
\exp[-i(t-t_0)\hat{H}_{_{SE}}]\hat{\rho}_{_{SE}}(t_0)
\exp[i(t-t_0)\hat{H}_{_{SE}}]$,
where  the initial state is
$\hat{\rho}_{_{SE}}(t_0)=\hat{\rho}_{_S}(t_0)\otimes\hat{\rho}_{_E}(t_0)$
and $\hbar=1$.
The {\it reduced} dynamics
 of the system $S$ is then defined as
\be
\hat{\rho}_{_S}(t) := \hat{\cal T}(t,t_0) \hat{\rho}_{_S}(t_0) =
{\rm Tr}_{_E} \left[\hat{\rho}_{_{SE}}(t)\right].
\label{4.1}
\ee
By definition, $\hat{\cal T}(t,t_0)$ is a linear map which transforms the
input state $\hat{\rho}_{_S}(t_0)$ onto the output state $\hat{\rho}_{_S}(t)$.
In our recent paper \cite{Buzek98}
 we have addressed the question {\it how to determine (reconstruct)
the master equation which governs the time evolution of the reduced
density operator $\hat{\rho}_{_S}(t)$}. It has been shown that
this master equation can be written in the {\it convolutionless}
 form (we omit the subscript $S$)
\be
\frac{d}{d t}\hat{\rho}(t) = \hat{\cal L}(t,t_0)\hat{\rho}(t).
\label{4.2}
\ee
which is possible due to the fact that in the {\it finite-dimensional}
Hilbert spaces matrix elements of density operators are analytic
functions. Consequently, $\hat{\cal T}(t,t_0)$ are non-singular
operators (except perhaps for a set of {\it isolated} values of $t$)
in which case the inverse operators $\hat{\cal T}(t,t_0)^{-1}$ exist
and the Liouvillian superoperator can be expressed as
\be
\hat{\cal L}(t,t_0):= \left[\frac{d}{dt}\hat{\cal T}(t,t_0)\right]
\hat{\cal T}^{-1}(t,t_0).
\label{4.3}
\ee
We note that $\hat{\cal T}(t,t_0)$   is uniquely specified by
 $\hat{H}_{_{SE}}$ and by the initial
state $\hat{\rho}_{_E}(t_0)$ of the environment.

\begin{figure}
\leftline{\epsfig{width=9cm,file=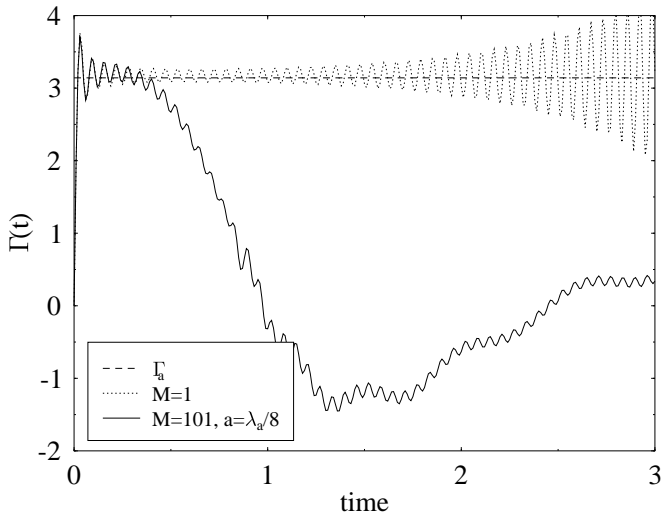}}
\caption{\narrowtext
The time evolution of the time-dependent
decay rate $\Gamma(t)$.
We assume the initially excited
atom to be in the center of the cavity,
other settings are as in  Fig.~\ref{fig1}.
In the chosen units we obtain from the Fermi golden role
(see \ref{decay1}) the decay rate  $\Gamma_a=\pi$ (see dashed line).
We consider two cases: firstly when there is just a single excited
atom in the cavity (dotted line) and secondly, when the excited atom
is surrounded by $100$ atoms (solid line)
 which create a ``linear'' crystal
with regular spacing between atoms ($a=\lambda_a/8$).
}
\label{fig8}
\end{figure}

In an earlier paper \cite{Buzek98} we have propose a general algorithm how
to reconstruct the Liouvillian superoperator
 $\hat{\cal L}(t,t_0)$  from
the knowledge of the unitary evolution
of the composite $S\oplus E$ system [see Eq.(\ref{4.1})]. From this
knowledge the master equation (\ref{4.2}) can be uniquely determined.
The dynamics of the open system (in our case the atom) is given exclusively
in terms of the system operators. Environmental degrees of freedom are
completely eliminated from the reduced dynamics. Nevertheless,
the state of the environment may change during the time evolution
due to the interaction with the system.
That is, there is no need to  employ the assumption
that the environment is a ``big'' reservoir which does not change under
 the action of the system.

To apply the formalism presented in Ref.~\cite{Buzek98} we remind ourselves
that the initial state of the atom-field system considered in the present
paper reads
\be
|\Psi(t_0)\r = |e\r_{1}|g\r_{\vec{j}}|0\r_{\vec{k}},
\label{4.4}
\ee
where $|e\r_{1}$ describes the excited state of the initially excited
atom, while $|g\r_{\vec{j}}:=|g\r_{2}\otimes\dots
\otimes|g\r_{M}$ describes the rest of the  $M$
 atoms which are initially in the ground
state. The vector $|0\r_{\vec{k}}$ denotes the vacuum of the cavity field.
Because the model Hamiltonian $\hat{H}_{tot}$ (\ref{htot})
is chosen so that the number of excitations in the system is an integral
of motion we can express the state vector of the atom field system at
time $t$ as
\be
|\Psi(t)\r &=& c_{1}(t)|e\r_{1}|g\r_{\vec{j}}|0\r_{\vec{k}}
\nonumber
\\
&+&\sum_{j=2}^M
c_{j}(t)|g\r_{1}|e_j\r_{\vec{j}}|0\r_{\vec{k}}
\label{4.5}
\\
&+&\sum_k
d_k(t)|g\r_{1}|g\r_{\vec{j}}|1_k\r_{\vec{k}},
\nonumber
\ee
where $|e_j\r_{\vec{j}}$ describes the state vector of a set of
$M-1$ atoms out of which the $j$-th atom is excited, while
$|1_k\r_{\vec{k}}$ describes the state of the cavity field with the
$k$-th mode in the Fock state $|1\r$ and all other modes in the vacuum
state. Using this notation we can express the master equation for the
originally excited atom as \cite{Buzek98}
\be
\frac{\partial}{\partial t}\hat{\rho}
& =& i\frac{\delta(t)}{2}\left[\hat{\rho},\hat{\sigma}_+\hat{\sigma}_-\right]
\label{4.6}
\\
&+& \frac{\Gamma(t)}{2}\left[2 \hat{\sigma}_-
\hat{\rho}\hat{\sigma}_+ - \hat{\sigma}_+\hat{\sigma}_-\hat{\rho}
- \hat{\rho} \hat{\sigma}_+\hat{\sigma}_-\right],
\nonumber
\ee
where the time-dependent decay rate $\Gamma(t)$ and the time-dependent
dynamical energy shift $\delta (t)$ can be expressed through the probability
amplitude $c_1(t)$ as
\be
\Gamma (t)= {\rm Re}\left[\eta(t)\right];
\qquad
\delta (t)= {\rm Im}\left[\eta(t)\right].
\label{4.7}
\ee
where
\be
\eta (t)= -2 \left[\frac{1}{c_1(t)}\frac{d\, c_1(t)}{d t}\right];
\label{4.8}
\ee
In general the parameter $\eta$ cannot be derived in an analytical form.
In Fig.~\ref{fig8} we present results of numerical evaluation.
We assume the initially excited
atom to be in the center of the cavity.
In the chosen units we obtain from the Fermi golden rule
[see Eq.(\ref{decay1}) the decay rate  $\Gamma_a=\pi$ (see dashed line).
We consider two cases: firstly when there is just a single excited
atom in the cavity (dotted line) and secondly, when the excited atom
is surrounded by $100$ atoms (solid line)
 which create a ``linear'' crystal
with regular spacing between atoms ($a=\lambda_a/8$).
In the case of the single atom $\Gamma(t)$ oscillates around the
value $\Gamma_a$.  The amplitudes of these oscillations
are relatively small till the recurrence time when it eventually takes
negative values (i.e. the atom starts to absorb energy from the field
- compare with Fig.~\ref{fig5}).
In the second case which mimics the decays of the atom in dielectrics
the time evolution of $\Gamma(t)$ is more complex.
At the initial instants  $\Gamma(t)$
oscillates around the value $\Gamma_a$, but then it rapidly decreases
and takes negative values - this is correlated with the absorption of
energy from the wave packets  reflected by surrounding atoms (see
Fig.~\ref{fig5}).

\section{Conclusions}
\label{sec8}

In this paper we have numerically studied a microscopic model
of the cavity QED describing an atom interacting with multimode
electromagnetic field. The initially excited
atom  is surrounded by a set of other atoms which represent a
dielectric ``crystal'' on a microscopic level. We have shown how the decay
of the atom is modified due to the mode structure of the cavity field
and the presence of neighboring atoms.

In the cavity QED model considered here
we have  neglected all mechanical effects
of the EM field on atoms. To account for effects of quantized
center--of--mass motion one can {\it generalize} the model
to a system of {\it trapped atoms} interacting with
EM field in the cavity.
We can assume a situation when due to laser cooling
the atoms embedded in a cavity (or PBS) are effectively trapped by
harmonic (optical) potential. There is a close analogy
with the model of trapped ions \cite{Buzek97b}.

In our next paper we  focus on the propagation of one-photon
wave packets in a cavity. These wave packets are represented
as superpositions of many cavity modes.
For example, we will study in detail
 scattering of a photon wave packet on
a single two-level atom. We will show that
 the atom within this framework effectively behaves as
 {\em quantum} beam splitter or semi-transparent mirror.
We will also discuss  dynamics of the wave packet incident
on a  ``crystal'' composed of two--level atoms which
fill some part of the resonator and thereby the transmission of the energy
of EM field depends on the density of the atoms.
This model will help us to understand on the microscopic level
how photon wave packets propagate through dielectric media and to
estimate (at least qualitatively) the speed of their propagation
from first principles. In addition the model will allow us to
formulate  the quantum version of the extinction theorem \cite{Buzek84}
which on the classical level \cite{Born80} explains how electromagnetic
waves packets propagate in material media.

\acknowledgements
We thank Rodney Loudon, Stig Stenholm, and Ed Hinds for helpful discussions.
This work was supported by the Slovak Academy of Sciences, the European
Union Network on Microlasers and Cavity Quantum Electrodynamics, the UK
Engineering and Physical Sciences Research Council and the Royal
Society.


\end{multicols}


\begin{thebibliography}{99}
\bibitem{footnote1}
We note here that the electromagnetic
mode itself is specified by the ``global'' structure of the cavity. This means
that even though the energy from the atom (local particle) to the field
is transferred	locally, the particular field mode becomes ``instantly''
excited in the whole cavity.
\bibitem{foot6}
We note that one can  create a
 quantum network in a regular
 optical lattice using  atoms from the Bose--Einstein
condensate (BEC).
\end{thebibliography}
\end{document}